\documentclass[%
 reprint,
superscriptaddress,
floatfix,
longbibliography,
 amsmath,amssymb,
 aps,
]{revtex4-2}

\usepackage{graphicx}
\usepackage{subfig}
\usepackage{dcolumn}
\usepackage{bm}
\usepackage{xurl} 
\PassOptionsToPackage{breaklinks}{hyperref}
\usepackage{hyperref}

\hypersetup{
    colorlinks=false,
    pdfborder={0 0 0},
}

\usepackage{amsmath}
\usepackage{caption}

\usepackage{xpatch}
\usepackage{xcolor}
\usepackage{float}

\hypersetup{
	colorlinks,
	linkcolor={black!100!black},
	citecolor={black!100!black},
	urlcolor={black!100!black}
}
\ProvideTextCommand{\DJ}{OT1}{\raisebox{0.25ex}{-}\kern-0.4em D}

\begin{document}

\preprint{APS/123-QED}

\title{A theoretical understanding of ionic current through a nanochannel driven by a viscosity gradient}

\author{Amer Alizadeh}
	\email {Corresponding author: amer.alizadeh@ucalgary.ca}
	\affiliation {Department of Chemical and Petroleum Engineering, Schulich School of Engineering,University of Calgary, AB,Canada.}%
\author {Hirofumi Daiguji}
	\affiliation {Department of Mechanical Engineering, The University of Tokyo, 7-3-1 Hongo, Bunkyo-ku, Tokyo 113-8656, Japan}
\author{Anne M. Benneker}
		\email {Corresponding author: anne.benneker@ucalgary.ca}
	\affiliation {Department of Chemical and Petroleum Engineering, Schulich School of Engineering,University of Calgary, AB,Canada.}%


\date{\today}

\begin{abstract}
It has been recently shown that a viscosity gradient could drive electrical current through a negatively charged nanochannel (Wiener and Stein, arXiv: 1807.09106). To understand the physics underlying this phenomenon, we employed the Maxwell-Stefan equation to obtain a relation between the flux of solvent species and the driving forces. Our 1D model, which was derived for both ideal and non-ideal solvents, shows that the ionic current depends on the ideality of the solvent, though both scenarios demonstrated good agreement with experimental data. We employed the model to understand the impact of solution bulk ionic strength and pH on the drift of ionic species with same reservoirs solution properties. Our modeling results unveiled the significant impact of bulk solution properties on the drift of ions which is in agreement with the experiments. Moreover, we have shown that the diffusion gradient along the nanochannel contributes significantly into driving ionic species if we even apply a small ionic concentration gradient to both reservoirs. Our modeling results may pave the way for finding novel applications for drift of ions toward a diffusion gradient.
\end{abstract}

\maketitle

\section{\label{sec:level1}Introduction}
Migration of ionic species is triggered by asymmetrical system conditions such as gradients of concentration \cite{RN118,RN178}, pressure \cite{RN139}, temperature \cite{RN1928,RN1825,RN1840}, and electrical potential \cite{RN1911}. These methods of ion transport phenomena have found numerous applications in energy harvesting \cite{RN488,RN186,RN461,RN1830}, water treatment \cite{RN176,RN385,RN121,RN357}, or logical part of nanofluidic chips \cite{RN283,RN491,RN715,RN1928,RN732}. Recently, Wiener and Stein \cite{wiener2018electrokinetic} have shown that by applying a viscosity gradient along a nanochannel one can drive an ionic current, similar to the other asymmetrical conditions. They deduced that this ionic current is a result of ions drifting toward the lower viscosity solvent, in which ions encounter a higher diffusivity. \par
Employing a viscosity gradient to drive particles has been studied for long time. For instance, Qiu and Mao \cite{RN1935} employed a viscosity gradient to separate different size nanoparticles. In their experiments, a pure viscosity gradient has been established by aqueous ployvinylpyrrolidone (PVP) solutions which could have considerable viscosity differences but have almost the same densities. They have shown that the viscosity gradient can be employed to effectively separate nanoparticles as an alternative to the density gradient-based methods. Hendrick de Haan and Salter \cite{RN1932} have theoretically shown that a polymer can translocate into a nanopore separating solvents of different viscosity. Their theory predicted that a pumping effect arises when the viscosity along the nanopore changes. Their modeling results stated that the direction of polymer translocation could be toward higher viscosity based on Brownian dynamics or, in contrast, toward lower viscosity by Langevin dynamics. Liebchen et al. \cite{RN1934} investigated the swimming of micro-particles such as bacteria toward favorable viscosity, so-called \textit{visco-taxis}. They have theoretically shown that the body shape of the biological or synthetic particles will create visco-taxis. Their model predicted that the shape of these micro-particles may prevent them to drift towards the lower viscosity region where they loose their swimming ability. Later, Datt and Elfring \cite{RN1933} developed a theory for the visco-taxis of active particles. They suggested that an effect, namely the impact of viscosity gradient on the thrust force generated by the micro-swimmers, which was not considered by Liebchen et al. \cite{RN1934} could determine the visco-taxis direction of these particles. Recently, Qiu et al. \cite{RN1936} found that a viscosity gradient applied to a nanopore can manipulate ionic rectification behavior. They reported the unexpected finding that by increasing the viscosity gradient, the rectification ratio increases for the forward bias applied electric field. Furthermore, Kurup and Basu \cite{Kurup2014ViscophoresisMA} studied the migration and sorting of droplets in a viscosity gradient, so-called \textit{visco-phoresis} which is introduced to a PDMS microchannel with different inlets and outlets. They have shown that if we have a parallel laminar stream of low and high viscosity, a droplet will migrate toward the lower viscosity by passing through the interface of the phases. They demonstrated that the viscosity gradient could separate the droplet by their sizes.   \par
Considering the above mentioned works, it is clear that different terminology was proposed (visco-taxis, visco-phoresis) for describing a particle movement inside a viscosity gradient. In this paper, we aim at a mechanistic understanding of the Wiener and Stein \cite{wiener2018electrokinetic} method in driving ionic species by employing a viscosity gradient. Since we are dealing with ionic species, we suggest \textit{visco-migration} as the terminology for drifting \textit{ions} in a viscosity gradient, analogous to electro-migration in which ions migrate under the influence of an external electric field. We will provide a theory based on the Maxwell-Stefan diffusion equations for ideal and non-ideal fluids. Successively, a simple equation will be developed to describe the relation of ionic drift velocity and the ionic current. For ideal fluids, Wiener and Stein \cite{wiener2018electrokinetic} developed a simple relation to obtain the diffusion and as a result the viscosity along the nanochannel by employing the Maxwell-Stefan equations. Their modeling results are in good agreement with their experimental measurements, but we extend the framework to a mechanistic approach which compared the ideal and non-ideal fluids with increasing the deviation of the fluid from the ideal scenario. After evaluating our model with the experimental data, we study the impact of bulk solution properties (i.e. pH and ionic strength) on the visco-migration ionic current. We obtained the zeta potential of the nanochannel by utilizing an EDL model, so-called electrical quad-layer model \cite{RN919}. Last but not least, we developed a simple equation to obtain the total ionic current when we not only apply a small ionic concentration gradient but also we have diffusion gradient owing to the viscosity gradient. 

\section{\label{sec:level2}Problem definition}

\begin{figure}[ht!]
    \centering
    \includegraphics[width=\columnwidth]{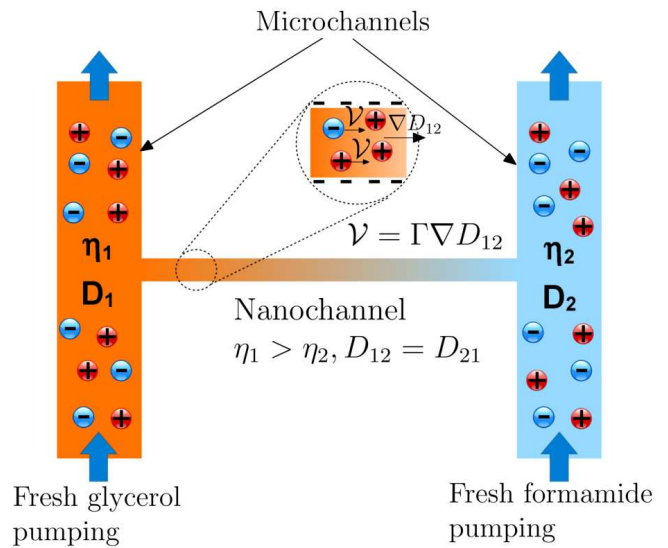}
    \caption{Schematic illustration of the nanochannel under viscosity gradients at both ends. It is assumed that the both ends of the nanochannel always subjected to fresh solutions with identical solution properties (i.e, bulk ionic strength and pH). The mixture of water \& glycerol and water \& formamide have different viscosity, $\eta_1$ and $\eta_2$, respectively, which results in different diffusion coefficients. The imposed diffusion gradient will drive ionic species toward higher diffusivity with velocity $\mathcal{V}$. Because of the nanochannel's counter-ion selectivity (overlapped electrical double layer), an ionic current will be measured.}
    \label{fig1}
    \centering
\end{figure}

Recently, Wiener and Stein \cite{wiener2018electrokinetic} have experimentally shown that a nanochannel bridging two microchannels filled with different viscosity solutions containing dissolved ionic species will drive an ionic current, even when the bulk ionic concentration in both fluids is equal. Fig. \ref{fig1} presents a 2D schematic illustration of the system which consists of two parallel microchannels and a nanochannel connecting them. It is assumed that the microchannels are continuously refreshed with glycerol and formamide solution allowing for the assumption that the viscosity gradient over the nanochannel is kept constant as a function of time. \par
In this contribution, we assumed that the mixtures of the glycerol \& water and formamide \& water are produced by considering the same volume fraction of water. We can obtain the water fraction for the mixtures by
 \begin{eqnarray}
 	\phi_{w,i}=\frac{\ln\left(\eta_{mix,i}\right)-\ln\left(\eta_{i}\right)}{\ln\left(\eta_w\right)-\ln\left(\eta_{i}\right)}
 	\label{eq1new}
 \end{eqnarray}
where $\mathit{\phi_{w,i}}$ denotes the water fraction in the mixture which is defined as $\mathit{\phi_{w,i}=c_{w,i}/\left(c_i+c_{w,i}\right)}$ and $\mathit{c_{w,i}}$ represents the concentration of water in this mixture. The volume fraction of glycerol/formamide for the left solvent mixture can be obtained as $\phi_i=1-\phi_{w,i}$. Given the volume fractions of water and glycerol/formamide, one can obtain the molar concentration as
 \begin{eqnarray}
 	C_i=\phi_i\times10^{3}\times\frac{\nu_i}{M_i} \label{eq2new}\\
 	C_{w,i}=\phi_{w,i}\times10^{3}\times\frac{1}{18.015}.
 	\label{eq3new}
 \end{eqnarray}

In all above equations, the subscript $i$ represents the glycerol ($g$) or formamide ($f$), $w$ stands for water, and $w,i$ for mixture of water \& glycerol/formamide. Finally, we can obtain the concentration (number of molecules/$\text{m}^3$) of each component as $c_i=N_A\times10^3\times C_i$ where $N_A$ is the Avogadro number. Table \ref{table1} summarizes the parameters that we used for this system. \par

\section{\label{sec:level3}Theory}

Generally, a system can be in thermodynamic or mechanical equilibrium. At thermodynamic equilibrium, the entropy production of the entire system is zero \cite{RN1047}. For a system at mechanical equilibrium, there is no acceleration of the center of mass while entropy production is non-zero \cite{RN1046}. For our problem, we assumed that the mechanical equilibrium is always accessible while the thermodynamic equilibrium is not reached. Thus, the entropy production can be defined as the production of flux and driving forces \cite{RN1029,RN1046, Holtan1953}
\begin{eqnarray}
T\sigma=J_qX_q+\sum_{i=1}^n\left(J_iX_i\right)
\label{eq7new}
\end{eqnarray} where $\sigma$, $\textit{T}$, $\textit{X}_q$, $\textit{X}_i$ and $J$ denote the entropy production strength, the temperature, thermodynamic force due to a temperature gradient, the species-driven thermodynamic forces and fluxes, respectively. The last term on the right-hand side of Eq. (\ref{eq7new}) is responsible for considering the total thermodynamic force owing to the flux of ionic species as well as the solvents, wherein $\textit{n}$ denotes the number of solution's components (i.e., $\textit{n}$=4 indicates two ionic species and two solvents). If we assume isothermal conditions, we can ignore the first term on the right hand side of Eq. (\ref{eq7new}) ($J_qX_q$). 
Regarding the thermodynamic force for driving the miscible liquids and ionic species we can write \cite{Holtan1953}
\begin{eqnarray}
	X_i=F_i-T\nabla\left(\frac{\mu_i}{T}\right)
	\label{eq8new}
\end{eqnarray}
where $\mathit{F_i}$ denotes the external body force (i.e. electrical body force) on the ionic and solvent species and $\mathit{\mu}_i$ represents the chemical potential of the $\textit{i}^{th}$ species.
If we extend the thermodynamic force for iso-thermal mass transfer (Eq. \ref{eq8new}) we will have
\begin{eqnarray}
	X_i=F_i-\left(\nabla{\mu_i}-\mu_i\frac{\nabla{T}}{T}\right).
	\label{eq9new}
\end{eqnarray}

Here, we should point out that we assume a zero effective external body force on the solvent molecules and ionic species. This assumption can be validated as the hydraulic resistance of the nanochannel is sufficiently high and balances the possible external pressure gradient. Consequently, the thermodynamic force which drives the solvent molecules and ionic species will be solely a function of the chemical potential gradient and could be written as
\begin{eqnarray}
	X_{i}=-\nabla{\mu_{i}}.
	\label{eq11new}
\end{eqnarray}
 
In the present work, we focus on the flux of solvents resulting from the chemical potential gradient at both ends of the nanochannel and we will investigate the related flux of ionic species which is a result of drifting toward diffusion gradient as well as an applied ionic concentration difference at both ends of nanochannel (c.f. section \ref{sec:level4}). Therefore, Eq. \ref{eq11new} can be expanded for solvent molecules as
\begin{eqnarray}
	X_i=-k_BT\nabla{\ln{c_i}},
	\label{eq12new}
\end{eqnarray}
where the index $\textit{i}$ changes from 1 to 2 to represent the glycerol \& water ($\textit{i}$=1) and formamide \& water ($\textit{i}$=2) mixture.  \par
In order to relate the species flux to the driving force, we start with the Maxwell-Stefan (MS) equation. The MS equation simply relates the friction forces between different components of a solution. In the mixture of the electrolyte solutions with different viscosity and in isothermal and isobaric conditions, the generalized MS equation is \cite{RN1029}
\begin{eqnarray}
	c_tk_BT\bold{d}_i=c_i\nabla_{T,p}{\mu_i},
	\label{eq13new}
\end{eqnarray}
where $\mathit{c_tk_BT\bold{d}_i}$ denotes the driving force on the solvent molecules which is expressed by \cite{RN1029}
\begin{eqnarray}
	\bold{d}_i=-\sum_{j=1,i\neq{j}}^n\frac{\phi_i \phi_j\left(\bold{u}_i-\bold{u}_j\right)}{D_{ij}},
	\label{eq14new}
\end{eqnarray}
and $\textit{n}$ is the number of solvents, $\mathit{\phi_i}$ and $\mathit{\phi_j}$ are the volume fraction of the species, defined as the ratio of the component concentration to the total concentration of the components in our system; $\mathit{\phi=c/c_t}$. $\mathit{D_{ij}}$ denotes the Maxwell-Stefan diffusion coefficient of species $\mathit{i}$ with respect to species $\mathit{j}$. We know that the species fluxes with respect to the solvent are defined as $\mathit{\bold{J}_i=c_i \bold{u}_i}$. If we introduce this relation into Eq. \ref{eq14new}, then we have
\begin{eqnarray}
	\bold{d}_i=\sum_{j=1,i\neq{j}}^n\frac{\phi_i \bold{J}_j-\phi_j \bold{J}_i}{c_tD_{ij}}.
	\label{eq15new}
\end{eqnarray}
where $\mathit{c_t}$ is the total number of solvent molecules per unit volume. Here it is worth pointing out that $\mathit{c_t}$ is constant along the nanochannel which is an essential assumption for an isobaric system. We can non-dimensionalize Eq. \ref{eq15new} by introducing $\mathit{\bar{x}=x/L},\mathit{\bar{c}=c/c_t},\mathit{\bar{D}_{ij}=D_{ij}/(\DJ_i+\DJ_j)}$, and $\mathit{\bar{\bold{J}}=\bold{J}/((\DJ_i+\DJ_j)c_t/L)}$, where $\DJ$ denotes the self diffusion coefficient, into this equation which gives rise to
\begin{eqnarray}
	\bold{\bar{d}}_i=\sum_{j=1,i\neq{j}}^n\frac{\phi_i \bold{\bar{J}}_j-\phi_j \bold{\bar{J}_i}}{\bar{D}_{ij}}.
	\label{eq16new}
\end{eqnarray}
where by considering Eq. \ref{eq13new}, we can relate $\mathit{\bold{\bar{d}}_i}$ to gradient of volume fraction of species $\mathit{i}$ as $\mathit{\bold{\bar{d}}_i=\sum_{j\text{=1}}^{\text{n-1}}\Gamma_{ij}\nabla\phi_i}$, where $\Gamma_{ij}$ represents the activity coefficient of species $\textit{i}$ in the the mixture and defined as \cite{RN1029}
\begin{eqnarray}
	\Gamma_{ij}=\delta_{ij}+\phi_i\frac{\partial{\ln\gamma_i}}{\partial{\phi_j}}.
	\label{eq17new}
\end{eqnarray}
For dilute gas mixtures, it is reasonable to assume the mixture as an ideal fluid where $\gamma_i=1$, however for higher density fluids or dense gases, $\gamma_i$ will be a function of the volume fractions. For instance, for a binary solution, we have
\begin{eqnarray}				
	\ln\gamma_1=A\phi_2^2,
	\label{eq18new}
\end{eqnarray}
which we will use for a non-ideal mixture scenario.
Hereinafter, for sake of avoiding more symbols, we drop all over-bars. Therefore, all parameters are non-dimensional, unless otherwise noted.
 
\subsection{Binary ideal mixture}
Considering an ideal binary system, Eq. \ref{eq16new} can be re-written as
\begin{eqnarray}
	\mathcal{X}_1=\frac{\phi_1\bold{J}_2-\phi_2\bold{J}_1}{D_{12}},
	\label{eq19new}
\end{eqnarray}
\begin{eqnarray}
	\mathcal{X}_2=\frac{\phi_2\bold{J}_1-\phi_1\bold{J}_2}{D_{21}},
	\label{eq20new}	
\end{eqnarray}
where $\mathcal{X}_i=\nabla{\phi_i}$ when $\Gamma_{ij}=1$. Eqs. \ref{eq19new} and \ref{eq20new} have four unknowns ($\phi_1,\phi_2,\bold{J}_1$, and $\bold{J}_2$) indicating we need two more equations to close the system of equations. Because of the Gibbs-Duhem restriction we have $\sum_{i=1}^{n}\mathcal{X}_i=0$. This gives rise to the symmetrical behavior of MS diffusion coefficient $D_{12}=D_{21}$ according to Eqs. \ref{eq19new} and Eq. \ref{eq20new}.\par
The isobaric condition enforces the total number of molecules along the nanochannel to be constant which requires that the sum of the volume fractions must be
 \begin{eqnarray}
 	\phi_1(x)+\phi_2(x)=1.
 	\label{eq21new}
 \end{eqnarray}
When introducing Eq. \ref{eq21new} into Eqs. \ref{eq19new} and \ref{eq20new} we have
\begin{eqnarray}
	\mathcal{X}_1=\frac{\phi_1}{D_{12}}\left(\bold{J}_1+\bold{J}_2\right)-\frac{\bold{J}_1}{D_{12}},
	\label{eq22new}
\end{eqnarray}
\begin{eqnarray}
	\mathcal{X}_2=\frac{\phi_2}{D_{12}}\left(\bold{J}_1+\bold{J}_2\right)-\frac{\bold{J}_2}{D_{12}}.
	\label{eq23new}
\end{eqnarray}
Solving Eqs. \ref{eq22new} and \ref{eq23new} for $\bold{J}_1+\bold{J}_2$ and substituting $\mathcal{X}_2$ with $-\mathcal{X}_1$ (Eq. \ref{eq24new}) and $\mathcal{X}_1$ with $-\mathcal{X}_2$ (Eq. \ref{eq25new}), we finally have
\begin{eqnarray}
	\frac{1}{\phi_1}\left(D_{12}\mathcal{X}_1+\bold{J}_1\right)=\bold{J}_1+\bold{J}_2.
	\label{eq24new}
\end{eqnarray}
\begin{eqnarray}
	\frac{1}{\phi_2}\left(D_{12}\mathcal{X}_2+\bold{J}_2\right)=\bold{J}_1+\bold{J}_2.
	\label{eq25new}
\end{eqnarray}
The final equation required to close our system can be obtained by considering the fact that the right-hand side of Eqs. \ref{eq24new} and \ref{eq25new} is a function of both component 1 and 2 while the left-hand side is only function of component 1 and 2, respectively. As a result, the only solution to Eq. \ref{eq24new} and \ref{eq25new} is $\bold{J}_1+\bold{J}_2=0$, which justifies the fact that in an $\mathit{n}$ component system, only $\mathit{n}-1$ fluxes are independent \cite{RN1029}. Considering the left-hand side of Eq. \ref{eq24new}, we have $\bold{J}_1=-D_{12}\mathcal{X}_1$ which denotes the Fick's diffusion law. The same procedure can be applied to Eq. \ref{eq25new} to obtain $\bold{J}_2=-D_{12}\mathcal{X}_2$. \par

To obtain the flux of solvents ($\bold{J}_1$ and $\bold{J}_2$) we need to find the MS diffusion coefficient ($D_{12}$). For binary systems which do not present a large deviation from ideality, the Maxwell-Stefan diffusion of a mixed solution can be directly related to the intradiffusion coefficients (Fick's diffusion) of the both components through the relation \cite{RN1855}
\begin{eqnarray}
	D_{12}=\phi_1\DJ_2+\phi_2\DJ_1
	\label{eq26new}
\end{eqnarray}
where $\mathit{\DJ_\text{1}}$ and $\mathit{\DJ_\text{2}}$ denote the diffusion of glycerol and formamide in de-ionized water and are equal to $\DJ_g$ and $\DJ_f$ in Table \ref{table1}. \par
According to the literature \cite{RN1909}, Eq. \ref{eq26new} is suitable if no viscosity data is available for the mixture of solutions. Another model which was proposed by Vignes \cite{RN1854} is recommended in general for concentrated binary liquids as
\begin{eqnarray}
	D_{12}=\DJ_{1}^{\phi_1}\DJ_{2}^{\phi_2}.
	\label{eq27new}
\end{eqnarray}

As Eq. \ref{eq26new} and \ref{eq27new} show, the diffusion of the mixed solution is a function of the solvents volume fractions ($\mathit{D_{12}=f(\phi_\text{1}(x),\phi_\text{2}(x))}$). In the present study, we employ Eq. \ref{eq27new} to obtain the MS diffusion coefficient. \par
Our system can be considered as a 1D problem since we do not have any flux perpendicular to the nanochannel walls. Therefore, according to mass conservation for a steady state condition without any chemical reactions, we have

\begin{eqnarray}
\frac{dJ_1}{dx}=0 \rightarrow -\ln\left(\frac{\DJ_1}{\DJ_2}\right)\mathcal{X}_1^2=\frac{d\mathcal{X}_1}{dx}
\label{eq28new}
\end{eqnarray}
Eq. \ref{eq28new} can be simplified by introducing $d\mathcal{X}_1/dx=\mathcal{X}_1d\mathcal{X}_1/d\phi_1$ and $\lambda=\ln(\DJ_1/\DJ_2)$ which gives rise to
\begin{eqnarray}
 -\lambda\mathcal{X}_1=\frac{d\mathcal{X}_1}{d\phi_1}, \mathcal{X}_1 \neq 0.
	\label{eq29new}
\end{eqnarray}
Eq. \ref{eq29new} can be solved for $\phi_1(x)$ as
\begin{eqnarray}
	\phi_1(x)=\frac{1}{\lambda}\ln\left[\lambda\left(C_1x+C_2\right)\right], \lambda \neq 0 	\label{eq30new} \\
	BC: \phi_1(0)=1, \phi_1(1)=0,
	\label{eq31new}
\end{eqnarray}
where the constant parameters ($C_1$ and $C_2$) can be obtained by introducing the above boundary conditions into Eq. \ref{eq30new} as
\begin{eqnarray}
	\begin{split}
		C_1=\frac{1-\exp(\lambda)}{\lambda},
		\\
		C_2=\frac{\exp(\lambda)}{\lambda},
	\end{split}
	\label{eq32new}
\end{eqnarray}
Re-calling Eq. \ref{eq21new}, we have $\phi_2(x)=1-\phi_1(x)$. Once we obtained $\phi_\text{1}(x)$ and $\phi_\text{2}(x)$, we already have $D_\text{12}(x)$ and, as a result, the flux of the solvents.

\subsection{Binary non-ideal mixture}
For a binary non-ideal mixture, we can write the MS equations as \cite{RN1029}
\begin{eqnarray}
	\Gamma\mathcal{X}_1=\frac{\phi_1\bold{J}_2-\phi_2\bold{J}_1}{D_{12}},
	\label{eq33new}
\end{eqnarray}
\begin{eqnarray}
	\Gamma\mathcal{X}_2=\frac{\phi_2\bold{J}_1-\phi_1\bold{J}_2}{D_{21}},
	\label{eq34new}	
\end{eqnarray}
where based on the Gibbs-Duhem restriction we have $\Gamma(\mathcal{X}_1+\mathcal{X}_2)=0$ and isobaric conditions imposes $\phi_1(x)+\phi_2(x)=1$. The activity coefficient for a binary mixture can be found by considering the Gibbs excess energy as $\mathcal{Q}=A\phi_1\phi_2$ which gives rise to \cite{RN1029}
\begin{eqnarray}
	\Gamma=1-2A\phi_1(x)\phi_2(x),
	\label{eq35new}
\end{eqnarray}
where $A$ is an arbitrary parameter to determine the deviation of the mixture from ideality.\par
If we introduce Eq. \ref{eq35new} into Eqs. \ref{eq33new} and \ref{eq34new} and follow the same approach as for the ideal mixture, we finally have
\begin{eqnarray}
	\frac{1}{\phi_1}\left(\Gamma D_{12}\mathcal{X}_1+\bold{J}_1\right)=\bold{J}_1+\bold{J}_2,
	\label{eq36new}
\end{eqnarray}
which gives rise to $\bold{J}_1=-\Gamma D_{12}\mathcal{X}_1$. Likewise the approach we employed to ideal mixture, we have
\begin{eqnarray}
	\frac{dJ_1}{d\phi_1}=0 \rightarrow 
	\left(\Gamma \ln(\DJ_1)-2A(\phi_2)\right)\mathcal{X}_1=-\Gamma \frac{d\mathcal{X}_1}{d\phi_1},
	\label{eq37new}
\end{eqnarray}
where for sake of having simpler form of ordinary differential equation, we employed the chain rule to the derivative of flux as
\begin{eqnarray}
	\frac{dJ_1}{dx}=\frac{dJ_1}{d\phi_1}\mathcal{X}_1=0
	\label{eq38new}
\end{eqnarray}
that for $\mathcal{X}_1 \neq 0$, we have $dJ_1/d\phi_1=0$.\par
Eq. \ref{eq37new} can be solved by separation of variables and taking integral of both sides which gives
\begin{eqnarray} \label{eq39new}
	\begin{aligned}
		&\ln(\mathcal{X}_1)= \\
		&-\lambda\phi_1-2A\int\frac{1-2\phi_1}{1-2A\phi_1\left(1-\phi_1\right)}d\phi_1+C_1^\prime,
	\end{aligned}
\end{eqnarray}
where by knowing that the integral has analytical solution as $\int\frac{1-2\phi_1}{1-2A\phi_1\left(1-\phi_1\right)}d\phi_1=\frac{\ln\left(2A\phi_1^2-2A\phi_1+1\right)}{2A}$, we re-arrange Eq. \ref{eq39new} by introducing $\mathcal{X}_1=d\phi_1/dx$ which gives rise to
\begin{eqnarray}
	\left(2A\phi_1^2-2A\phi_1+1\right)\exp\left(\lambda\phi_1\right)d\phi_1=C_1^\prime dx.
	\label{eq40new}
\end{eqnarray}
Integrating both sides of Eq. \ref{eq40new}, we finally have
\begin{eqnarray} \label{eq41new}
	\begin{aligned}
		&\frac{\mathrm{e}^{\lambda\phi_1(x)}}{\lambda}\left[2A(\phi_1(x)^2-\phi_1(x))+\frac{2A}{\lambda}(1-2\phi_1(x))+ \right. \\
		& \left. \frac{4A}{\lambda^2}+1\right]=C_1^\prime x+C_2^\prime, \lambda \neq 0.
	\end{aligned}
\end{eqnarray}
\begin{eqnarray}
	BC: \phi_1(0)=1, \phi_1(1)=0
	\label{eq42new}
\end{eqnarray}
Eq. \ref{eq41new} is a transcendental equation which has no closed-form solution for $\phi_1(x)$. Therefore, we can solve it by approximating the left hand-side by a Taylor expansion that for $\mathcal{O}(\phi_1(x)^{11})$ the maximum error will be on the of order $10^{-6}$. Consequently, we can safely utilize the Taylor expansion (around $x=0$) and solve Eq. \ref{eq41new} for $\phi_1(x)$ with the proper boundary conditions (Eq. \ref{eq42new}). It is worth pointing out that for $A=0$ (ideal fluid), Eq. \ref{eq41new} re-covers Eq. \ref{eq30new}.
 \begin{table}[h!]
	\caption{\label{tab:table1} Parameters employed to model the experimental measurements. Note that subscript $g$ and $f$ refer to glycerol and formamide, respectively}
	\begin{ruledtabular}
		\begin{tabular}{lcdr}
			\textrm{Parameters}&
			\textrm{Amounts}\\
			\colrule
			$\rho$  & 999.99 (kg/$m^3$)\\
			$\eta_{mix,g}$  &  3.4$\times10^{-3}$ (Pa.s)\\
			$\eta_{mix,f}$  & $10^{-3}$ to $5\times10^{-3}$ (Pa.s) \\
			$\eta_{w}$  & $8.9\times10^{-4}$ (Pa.s)\\
			$\eta_{g}$  & 0.934 (Pa.s)\\
			$\eta_{f}$  & $3.34\times10^{-3}$ (Pa.s)\\
			$M_g$  & 92.094 (g/mol)\\
			$M_f$  & 45.04 (g/mol)\\
			$\nu_{g}$   & 1.26 (g/$cm^3$)\\
			$\nu_{f}$   & 1.13 (g/$cm^3$)\\
			$\DJ_g=\DJ_1$	&2.2$\times10^{-12}$ ($m^2$/s) \\
			$\DJ_f=\DJ_2$	&5.5$\times10^{-10}$ ($m^2$/s) \\
			Channel length$\equiv$L  & 200 ($\mu$m)\\
			Channel Width$\equiv$W	& 150 ($\mu$m)\\
			Channel Height$\equiv$H	& 50 ($\mu$m)\\
			Vacuum permittivity$\equiv$$\epsilon_0$	& 8.854$\times$10$^{-12}$ (Fm$^{-1}$) \\
			Water electrical permittivity$\equiv$ $\epsilon_r$	& 78.54\\
			Boltzmann constant$\equiv$ $k_b$	& 1.38$\times$10$^{-23}$ (JK$^{-1}$)\\
			Solution temperature$\equiv$T	& 298.15 K\\
			Electron charge$\equiv$ $e$	&1.602$\times$10$^{-19}$ (C) 
		\end{tabular}
		\label{table1}
	\end{ruledtabular}
\end{table}

\section{\label{sec:level4}Results and Discussion}
So far, we found the volume fraction of the species along the nanochannel (Eqs. \ref{eq30new} and \ref{eq41new}) which provides the MS diffusion coefficient and species flux. In the following sections we will first benchmark our model and, second, investigate the impact of bulk solution \textit{p}H and ionic strength on the nanochannel's surface charge by solving the available electrical double layer models such as electrical-quad layer model \cite{RN919}. Finally, we will compare visco-migration and reverse electrodialysis ionic current by not only applying a viscosity gradient but also an ionic concentration gradient.
\subsection{Model benchmark}
We have solved Eqs. \ref{eq30new} and \ref{eq41new} to obtain the MS diffusion coefficient ($D_{12}(x)$) and successively the flux of species. As Wiener and Stein \cite{wiener2018electrokinetic} mentioned, their experimental data are consistent with the isothermal rule \cite{RN1930} in stochastic displacement models. According to the Hanggi model \cite{RN1930}, a particle drifts toward the lower viscosity (higher diffusion) with a speed
\begin{eqnarray}
	\mathcal{V}=\frac{dD_{12}(x)}{dx}
	\label{eq43new}
\end{eqnarray}
where, as we stated above, $D_{12}$ and $x$ are non-dimensional parameters. \par
Non-dimensionalizing the equations in section \ref{sec:level3} allowed us to figure out that for the ideal Eq. \ref{eq48new} and non-ideal Eq. \ref{eq49new} scenarios we have
\begin{eqnarray}
	\text{ideal:}\:{{J}_1}=\frac{-1}{\lambda}{\mathcal{V}}.
	\label{eq48new}
\end{eqnarray}
\begin{eqnarray}
	\text{non-ideal:}\:{J}_1=\frac{-\Gamma}{\lambda}{\mathcal{V}}.
	\label{eq49new}
\end{eqnarray}

Studying Eq. \ref{eq49new} reveals that since $\Gamma$ is a function of $x$ (Eq. \ref{eq35new}), therefore ${\mathcal{V}}$ for non-ideal fluids cannot be a constant amount in contrast to what we had for ideal fluids. In other words, since the left-hand side of Eq. \ref{eq49new} is constant amount (flux of species), as a result, the multiplication of $\Gamma\:{\mathcal{V}}$ must be a constant amount. This fact suggests that the Hanggi model (Eq. \ref{eq43new}) can be modified for ideal and non-ideal fluids as
\begin{eqnarray}
	{\mathcal{V}}=\Gamma\frac{d{D}_{12}(x)}{d{x}}.
	\label{eq50neww}
\end{eqnarray}
Therefore, we can simply derive a general equation for the flux of ideal and non-ideal species as function of drift velocity
\begin{eqnarray}
	J_1=\frac{-1}{\lambda}\mathcal{V}.
	\label{eq47neww}
\end{eqnarray}
Hereinafter, we will present the non-dimensional parameters with over-bar. Ionic species are typically moving in an isothermal domain owing to the three transport phenomena: advection, diffusion, and electromigration \cite{RN1911,RN111}. Since in Wiener and Stein's experiments \cite{wiener2018electrokinetic} there is no applied external electric field, pressure gradient, and ionic concentration gradient, we can simply relate the drift velocity to the advection of ionic species which gives rise to
\begin{eqnarray}
	I_{visco}=\left(\frac{\DJ_1+\DJ_2}{L}\right)\bar{\mathcal{V}}\: \rho_{ave}\: \mathcal{A}
	\label{eq44new}
\end{eqnarray}
where $\mathcal{A}=W \: H$ represents the cross sectional area of the nanochannel and the average net electric charge density $\rho_{ave}$ (Cm$^{-3}$) defined as
\begin{eqnarray}
	\rho_{ave}=\frac{\int_{0}^{H/2}\rho_edy}{H/2}.
	\label{eq45new}
\end{eqnarray}
In Eq. \ref{eq45new}, $\rho_e$ denotes the local net electric charge density which is obtained as
\begin{eqnarray}
	\rho_e=-\epsilon_0\epsilon_r\frac{d\psi(y)}{dy}
	\label{eq46new}
\end{eqnarray}
where $\psi(y)$ is the internal electric potential [V] owing to the electrical double layer effect which for a channel with similar charge on the walls can be obtained as \cite{RN111}
\begin{eqnarray}
	\psi(y)=\frac{\sigma}{\epsilon_0\epsilon_r\sinh{\left(\kappa H/2\right)}}\cosh\left(\kappa y\right).
	\label{eq47new}
\end{eqnarray}
It is worth noting that in Eq. \ref{eq47new}, $\sigma$ represents the measured surface charge on the nanochannel's wall which Wiener and Stein \cite{wiener2018electrokinetic} measured for their setup (100 mM NaCl) is $\sigma=-200$ (mC/m$^2$). Furthermore, $\kappa = \sqrt{\epsilon_0\epsilon_r k_b T/2e^2n_b}$ is the inverse of Debye length (m$^{-1}$).\par
Now, let us re-write Eq. \ref{eq44new} by substituting the drift velocity with the species flux from Eq. \ref{eq47neww} which reads
\begin{eqnarray}
	{I}_{visco}=-\left(\frac{\DJ_1+\DJ_2}{L}\right)\:\lambda\: {\bar{J}_1}\: {\rho}_{ave}\: {\mathcal{A}}.
	\label{eq50new}
\end{eqnarray}

Eq. \ref{eq50new} shows that the ionic current because of the visco-migration phenomenon can be obtained by multiplying the species fluxes by $\lambda$, average net charge density, and the cross-section of the nanochannel.

\begin{figure}[ht!]
	\centering
	\includegraphics[width=\columnwidth]{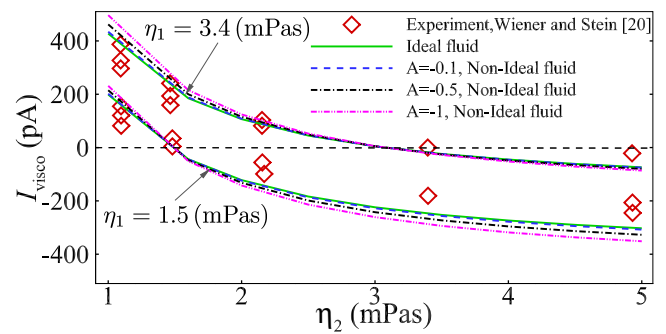}
	\caption{The ionic current $\left(I_{visco}\right)$ as a result of an applied viscosity gradient along the fabricated nanochannel by Wiener and Stein \cite{wiener2018electrokinetic} for two different glycerol \& water viscosities $(\eta_1=3.4$ and 1.5 [mPas]). Our ideal and non-ideal modeling results are validated with the experimental data.}
	\label{fig2}
	\centering
\end{figure}

Figure \ref{fig2} illustrates our ideal and non-ideal models' prediction against the experimental data for glycerol \& water with $\eta_1=3.4$ [mPas] and 1.5 [mPas]. Our modeling results show that the ionic current owing to visco-migration phenomenon is a function of the applied viscosity gradient as well as the deviation from ideal fluid. It has been unveiled that by increasing the $\eta_\text{2}$, the impact of deviation from ideality increases.  \par

\begin{figure*}[ht!]
	\centering
	\subfloat{\includegraphics[width=7cm]{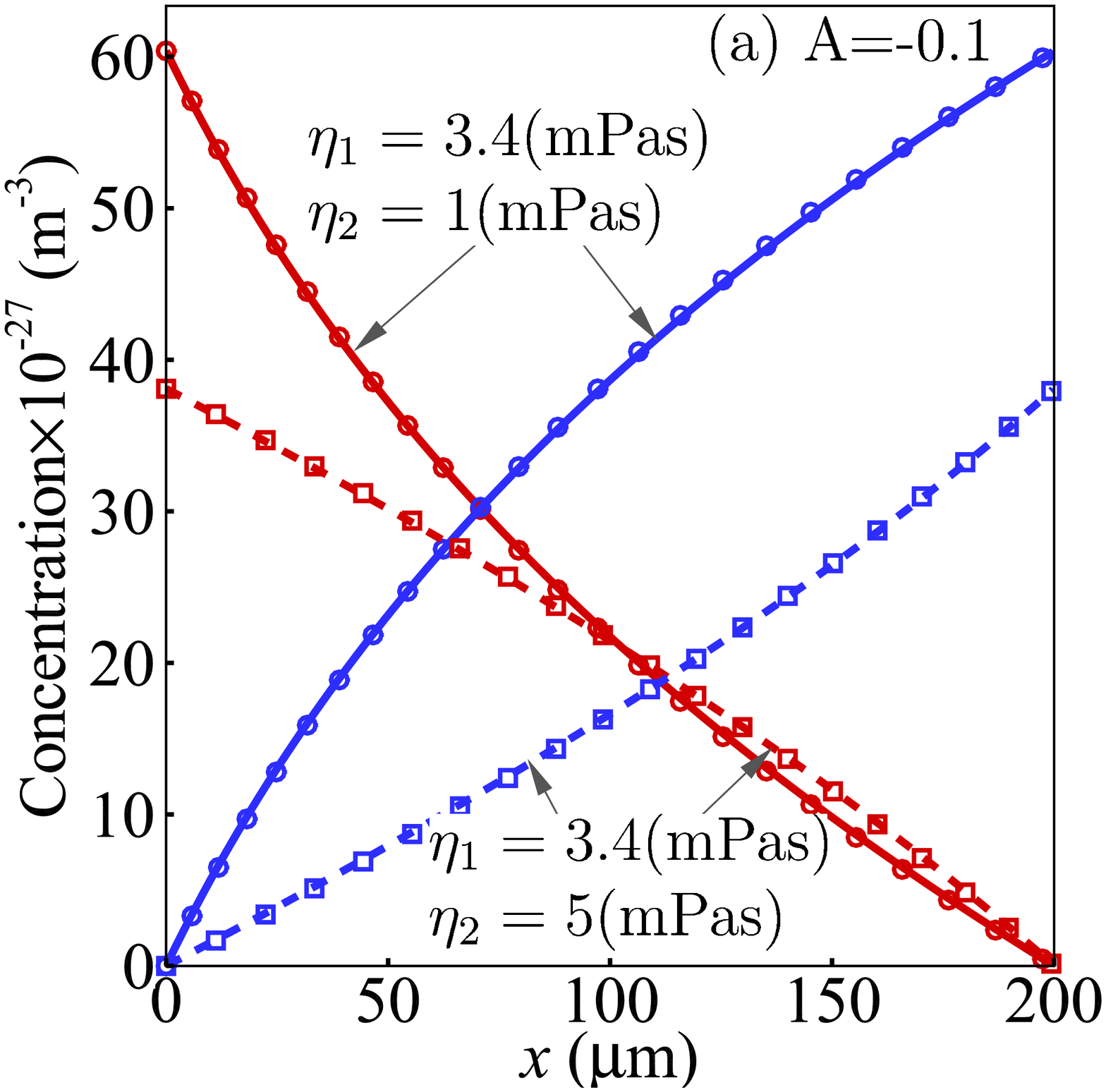}}
	\qquad
	\subfloat{\includegraphics[width=7cm]{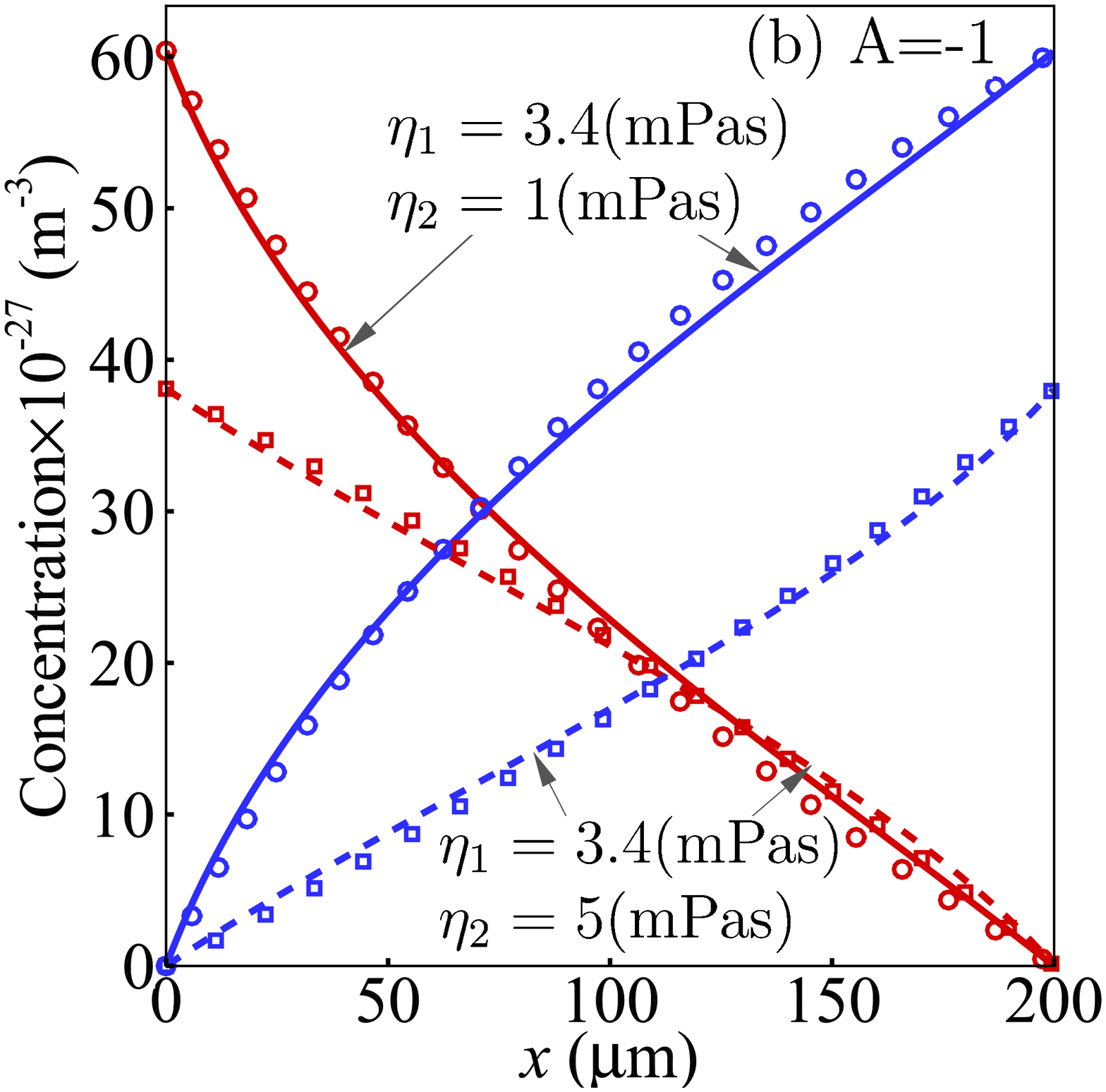}}
	\caption{Concentration of species 1 (red lines/symbols, glycerol \& water) and 2 (blue lines/symbols, formamide+water) along the nanochannel for different applied viscosity gradients. Symbols represent the ideal fluid ($\Gamma=1$) and non-ideal fluids are illustrated by solid and dashed lines for (a) $A=-0.1$ and (b) $A=-1$. }
	\label{fig3}
	\centering
\end{figure*}

After comparing the predicted ionic current $I_{visco}$ with the experimental measurements, we further scrutinize the distribution of species and the MS diffusion coefficient along the nanochannel. To this end, we present the concentration of species for ideal and non-ideal fluids under different viscosity gradients (Fig. \ref{fig3}) where the symbols represent the ideal and solid/dashed lines the non-ideal fluids. It is worth mentioning that for the present system, the red lines and symbols represent the concentration of glycerol \& water ($c_1$) while the blue ones stand for formamide \& water($c_2$). It can be seen that for $A=-0.1$ (Fig. \ref{fig3}a), the ideal and non-ideal models predict almost the same distribution of solvent species.  \par
\begin{figure*}[ht!]
	\centering
	\subfloat{\includegraphics[width=7cm]{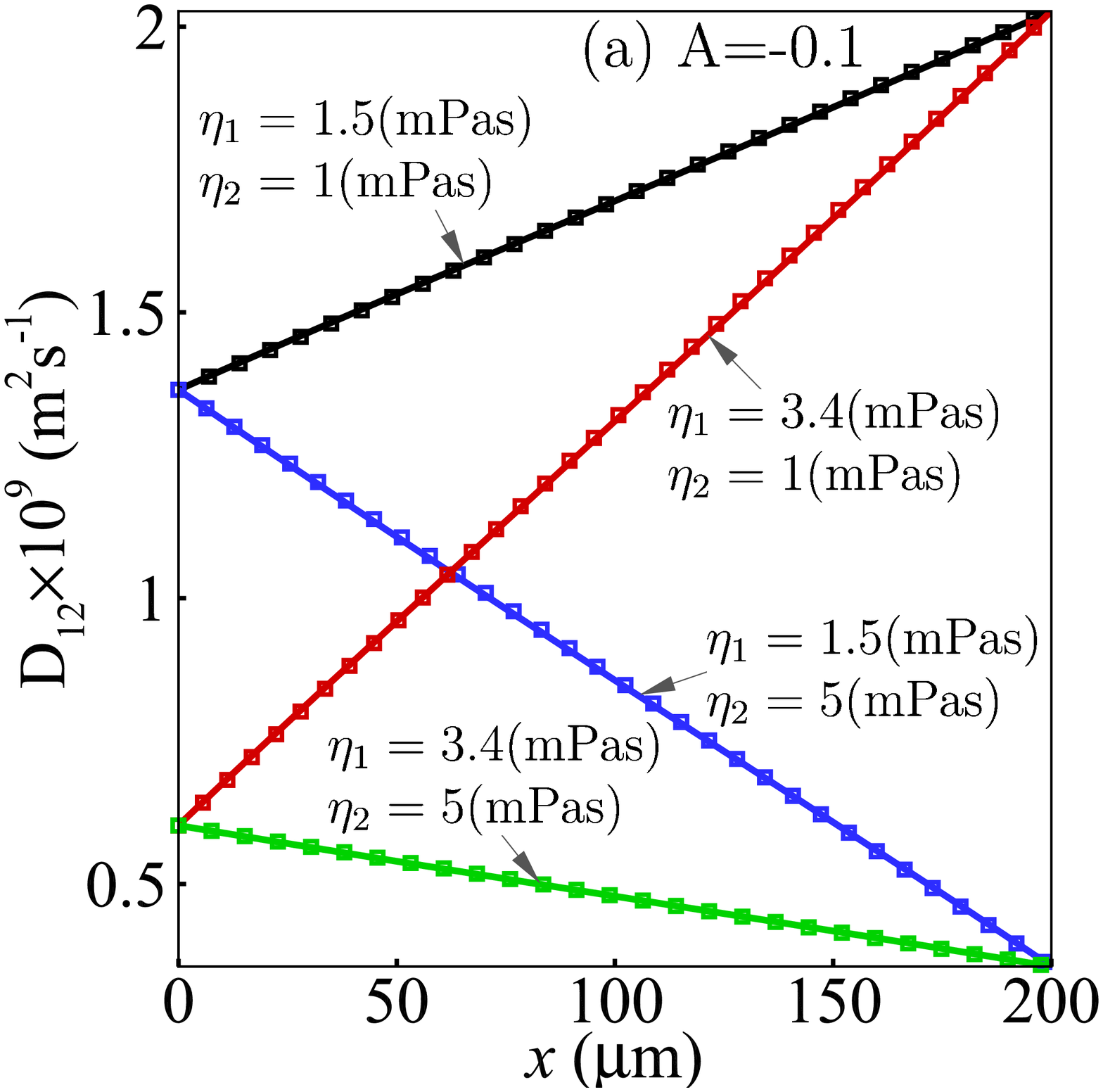}}
	\qquad
	\subfloat{\includegraphics[width=7cm]{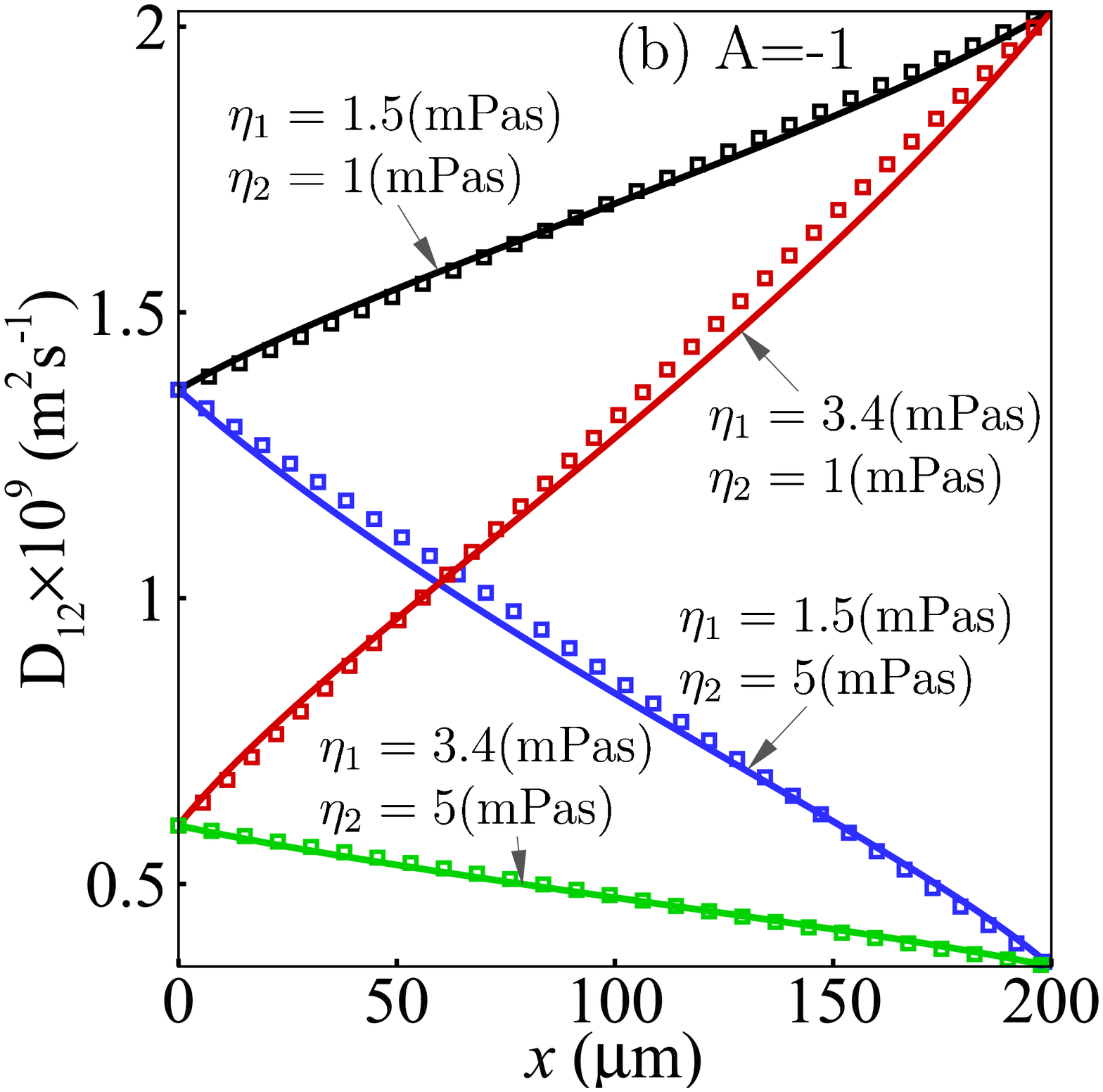}}
	\caption{The MS diffusion coefficient ($D_{12}$) versus the length of the nanochannel for ideal fluid ($\Gamma=1$) represented by the symbols and non-ideal fluid which demonstrated by the solid and dashed line for (a) $A=-0.1$ and (b) $A=-1$. }
	\label{fig4}
	\centering
\end{figure*}

By increasing $A$ from $-$0.1 to $-$1, Fig. \ref{fig3}b demonstrates that the ideal and non-ideal modeling results deviates, albeit rather small. As was the case for $A=-0.1$, for $A=-1$ we can see that by increasing the viscosity difference, an increasingly non-linear behavior of the species concentration as function of distance is developed. Another parameter that we are interested to see the behavior of it for different viscosity gradient and fluid ideality, is the MS diffusion coefficient. Fig. \ref{fig4} demonstrates $D_{12}$ as function of distance $x$. It is interesting to note that $D_{12}$ shows linear behavior versus the distance from the inlet of the nanochannel for both ideal fluid and non-ideal fluid with $A=-0.1$ (Fig. \ref{fig4}a). On the other hand, more deviation from the ideal fluid ,$A=-1$, results in non-linear behavior for $D_{12}(x)$.

\subsection{Impact of bulk pH and ionic strength}
\begin{figure*}[ht!]
	\centering
	\subfloat{\includegraphics[width=7cm]{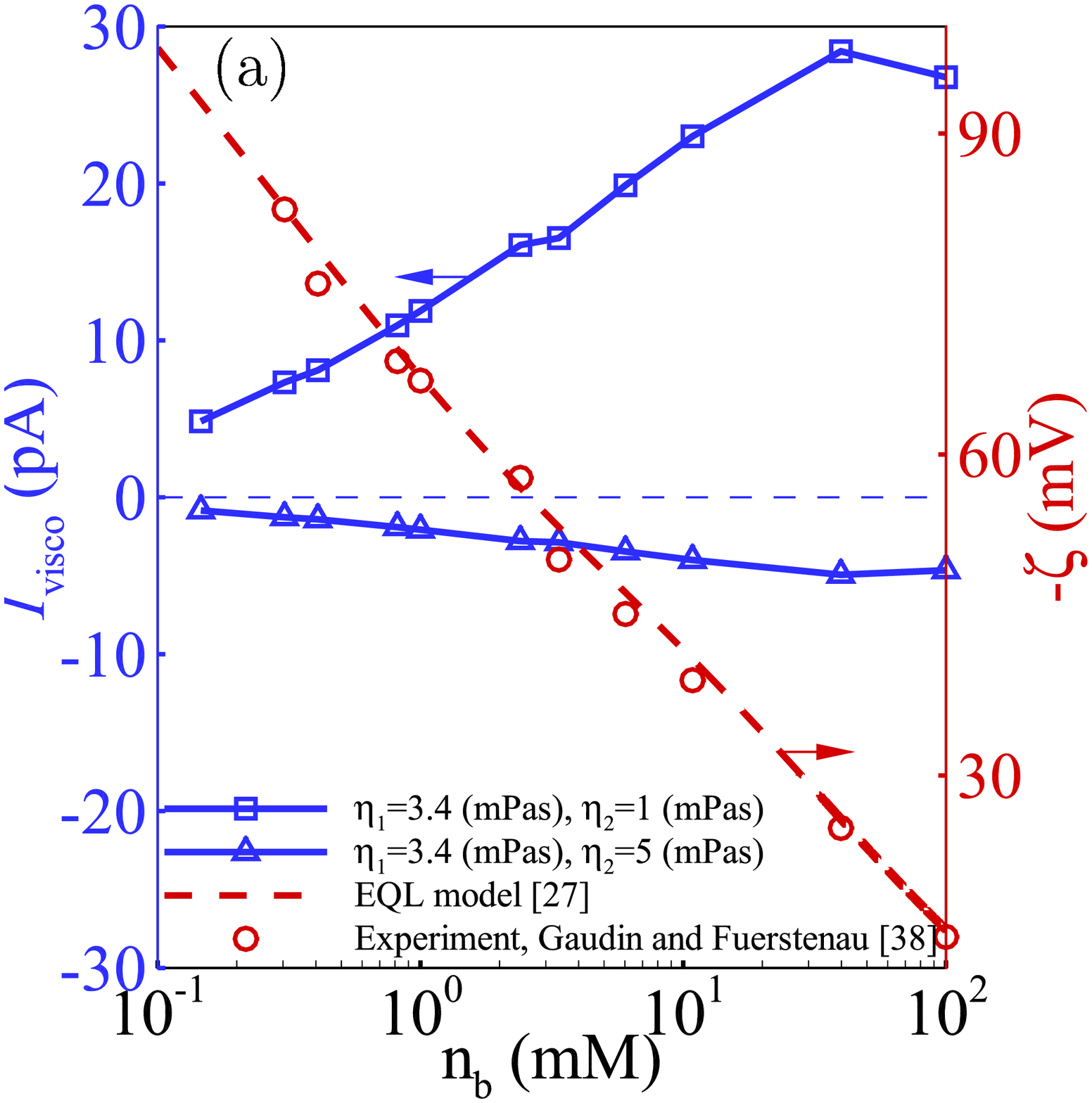}}
	\qquad
	\subfloat{\includegraphics[width=7cm]{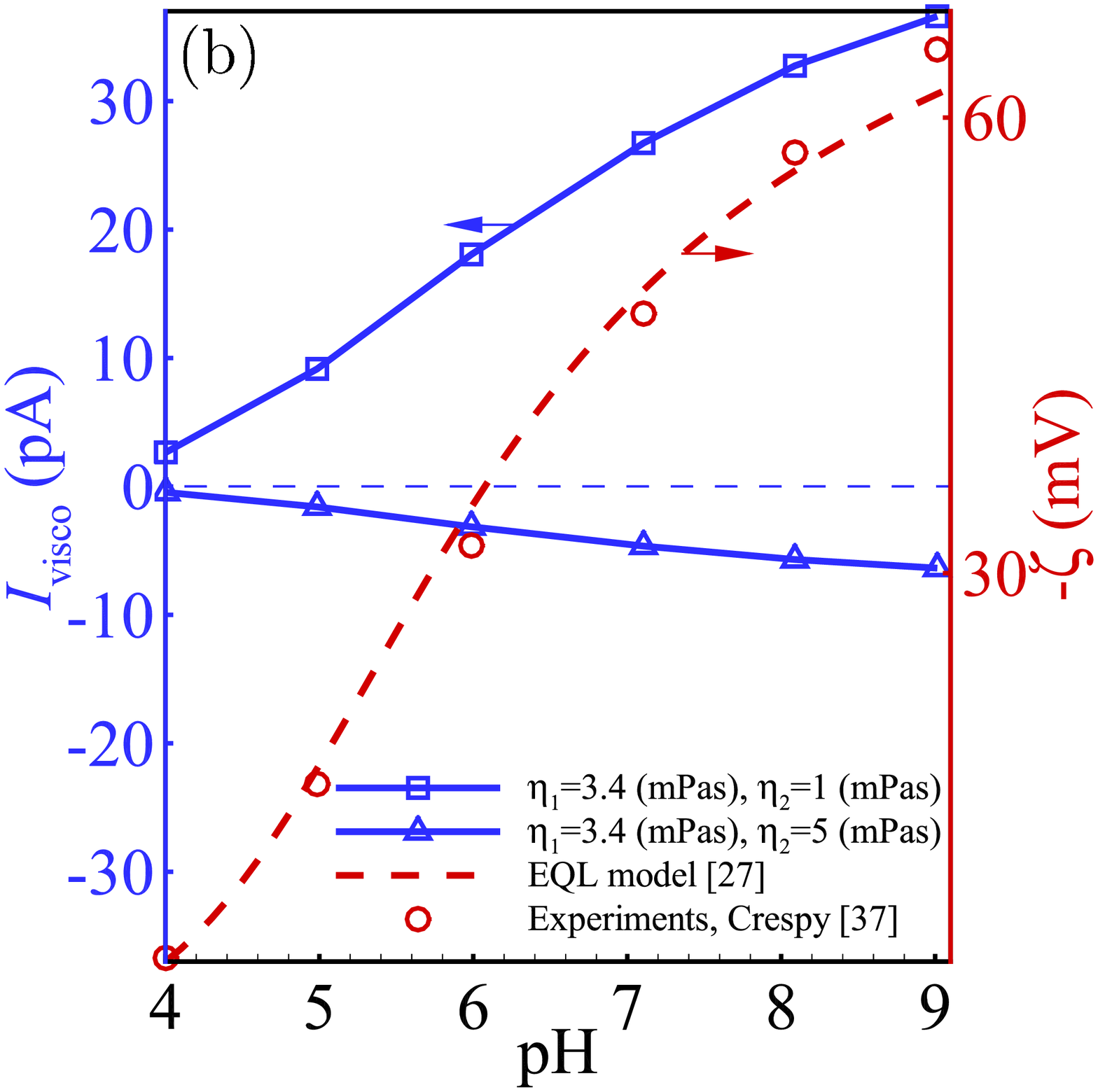}}
	\caption{The impact of (a) bulk ionic strength ($n_b$) with \textit{p}H=$6.5$ and (b) solution's \textit{p}H with $n_b=0.01$ M on $I_{visco}$ for NaCl solution. The zeta potentials to obtain the average net-charge density is shown by red line obtained by the EQL model \cite{RN919} and symbols the experimental measurements \cite{RN524}.}
	\label{fig5}
	\centering
\end{figure*}

In this section, we investigate the impact of bulk solution properties such as pH and bulk ionic concentration ($n_b$) on the resulting current $I_{visco}$ (Eq. \ref{eq50new}). To this end, we employ a recently developed electrical double layer, so-called electrical-quad layer model (EQL) \cite{RN919} by considering the following chemical reactions

\begin{eqnarray}
	\text{SiOH}^+_2\rightleftharpoons \text{SiOH}+\text{H}^+, K^{a1}_{int} \label{eq53new} \\
	\text{SiOH} \rightleftharpoons \text{SiO}^-+\text{H}^+, K^{a2}_{int} \label{eq54new} \\
	\text{SiO}^-+\text{M}^+ \rightleftharpoons \text{SiOM}, K^{a3}_{int} \label{eq55new}	
\end{eqnarray}
where M$^+$ denotes the counter-ion (i.e. Na$^+$), and $K^{a1}_{int}$ to $K^{a3}_{int}$ represent the chemical equilibrium constants. 
Solving the equations of the EQL model \cite{RN919} predicts the $\zeta$-potential of the silica-aqueous solution interface as function of bulk ionic strength (Fig. \ref{fig5}a) and the solution \textit{p}H (Fig. \ref{fig5}b) which were evaluated by the available experimental data \cite{RN372,RN524}. Since we are dealing with the $\zeta$-potential, the electrical potential distribution perpendicular to the nanochannel walls can be obtained by \cite{RN111}
\begin{eqnarray}
	\psi\left(y\right)=\frac{\zeta \cosh\left(\kappa y\right)}{\cosh\left(\kappa H/2\right)},
	\label{eq56new}
\end{eqnarray}

where $\zeta$ represents the zeta potential on the nanochannel walls. Here, we have to point out that the bulk reservoirs at both ends of the nanochannel have the same solution properties. Fig. \ref{fig5}a shows the $I_{visco}$ versus the bulk ionic strength for an NaCl solution with $\textit{p}\text{H}$=6.5 and for two different viscosity gradient scenario $\eta_1=3.4 \text{(mPas)}$, $\eta_2=1 \text{(mPas)}$ and $\eta_1=3.4 \text{(mPas)}$, $\eta_2=5 \text{(mPas)}$. Our modeling results show that by increasing the bulk ionic concentration the absolute value of $I_{visco}$ increases to a maximum amount and then decreases by further increasing $n_b$. This behavior of the ionic current can be attributed to the competition of the EDL thickness and $\rho_{ave}$, where by increasing $n_b$, the absolute value of $\rho_{ave}$ increases and then decreases. In other words, when we increase the bulk ionic concentration, the surface charge density at the nanochannel walls increases \cite{RN1911,RN919,RN203}. However, for higher amount of $n_b$, leading to thinner electrical double layers, the major part of the nanochannel is electro-neutral ($\rho_e=0$) which does not contribute into the average net charge density.
 Moreover Fig. \ref{fig5}a illustrates that the $I_{visco}$ has higher sensitivity to the bulk ionic concentration when we have $\eta_1=3.4 \text{(mPas)}$, $\eta_2=1 \text{(mPas)}$ compared to $\eta_1=3.4 \text{(mPas)}$, $\eta_2=5 \text{(mPas)}$. \par
Regarding the impact of solution $\textit{p}$H, Fig. \ref{fig5}b shows that for an NaCl solution with $n_b=0.01 \text{M}$, the absolute value of $I_{visco}$ always increases with pH while similar to Fig. \ref{fig5}a, $I_{visco}$ increases more for viscosity gradient $\eta_1=3.4\text{(mPas)}$, $\eta_2=1\text{(mPas)}$. Comparing Fig. \ref{fig5}a and \ref{fig5}b shows that although higher \textit{p}H provides a smaller absolute value of zeta potential in comparison with the bulk ionic strength, the $I_\text{visco}$ for both scenarios are almost in the same order of magnitude. 
\subsection{Visco-migration vs Reverse Electrodialysis}
So far we have investigated the impact of solution properties on the ionic current owing to the viscosity gradient. As we have previously alluded to, for $I_\text{visco}$, the sole existence of the ionic current is a result of the drift of ionic species because of the diffusion gradient resulting from the viscosity gradient. However, since the viscosity of the solvent is changing along the nanochannel, based on the Stokes-Einstein equation, the diffusion of the ionic species must be changed which suggests that if we apply a small concentration gradient of ionic species, one can take the advantage of this diffusion gradient to drive an increasing ionic current \cite{RN337,RN461,RN487}. In this section, we aim at proposing a simple analytical solution to the Nernst-Planck equation when we only have ionic species concentration gradient as
\begin{eqnarray}
	j_i=-D_i\nabla c_i
	\label{eq57new}
\end{eqnarray}
where $j_i$ denotes the flux, $D_i$ the diffusion, and $c_i$ the concentration of the $i$th ionic species. It is worth noting that in Eq. \ref{eq57new} we dropped the electro-migration term since we assumed that the applied concentration gradient is small enough to ignore the surface charge heterogeneity and no external electric field is applied \cite{RN873}.

\begin{figure}[ht!]
	\centering
	\includegraphics[width=\columnwidth]{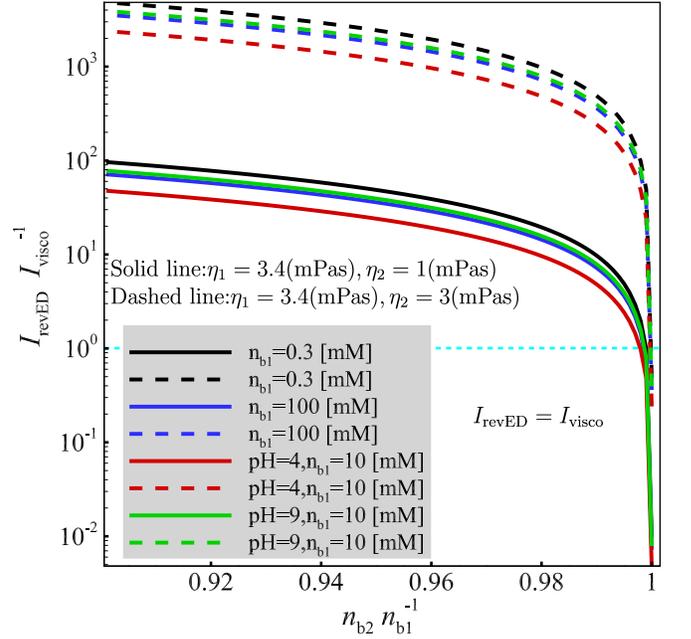}
	\caption{The relation of ionic current because of concentration gradient ($I_\text{revED}$) to ionic current owing to viscosity gradient ($I_\text{visco}$) versus the deviation of right reservoir bulk ionic concentration from the left reservoir. Dashed and solid lines represent different viscosity gradients for two bulk ionic concentration and solution pH.}
	\label{fig7}
	\centering
\end{figure}

Let us define $n_s=c_1-c_2$ and assume that the diffusivity of both co- and counter-ions are approximately identical ($D_1\simeq D_2=D^\circ$). Now we can subtract the flux of the counter-ion ($i$=1) from co-ion ($i$=2) based on Eq. \ref{eq57new} which gives rise to
\begin{eqnarray}
	j_\text{net}=-D^\circ \nabla n_s.
	\label{eq58new}
\end{eqnarray}
Employing the mass conservation law to Eq. \ref{eq58new} gives
\begin{eqnarray}
	B\frac{d\langle n_s \rangle}{dx}+\left(A+Bx\right)\frac{d^2 \langle n_s \rangle}{dx^2}=0
	\label{eq59.5new}
\end{eqnarray}
where $B=dD^\circ/dx$. Eq. \ref{eq59.5new} can be solved analytically (see appendix A for definition of $A$ and $B$) as
\begin{eqnarray}
	\langle n_s \rangle =a_1+a_2\ln\left(x+\frac{A}{B}\right)
	\label{eq59new}
\end{eqnarray}
where $\langle n_s \rangle=\left(1/H/2\right)\int_{0}^{H/2}n_sdy$ represents the average amount of $n_s$ perpendicular to the nanochannel's walls. Eq. \ref{eq59new} can be solved by having the boundary conditions
\begin{eqnarray}
	x=0 : \langle n_{s1} \rangle=-2n_{b1}\frac{\int_{0}^{H/2}\sinh\left(\frac{\psi\left(y\right)}{V_T}\right)dy}{H/2} \label{eq60new} \\
	x=L : \langle n_{s2} \rangle=-2n_{b2}\frac{\int_{0}^{H/2}\sinh\left(\frac{\psi\left(y\right)}{V_T}\right)dy}{H/2} \label{eq61new}
\end{eqnarray}
where $n_{b1}$ and $n_{b2}$ are the bulk ionic concentration at the left and right reservoirs, respectively, and $V_T$ represents the thermal voltage which is defined as $k_bT/e$. Introducing Eqs. \ref{eq60new} and \ref{eq61new} into Eq. \ref{eq59new}, we have
\begin{eqnarray}
	\begin{split}
		a_1=\langle n_{s1} \rangle-a_2\ln\frac{A}{B} \\
		a_2=\frac{\langle n_{s2} \rangle - \langle n_{s1} \rangle}{\ln\left(1+\frac{B}{A}\right)}.
	\end{split}
	\label{eq63new}
\end{eqnarray}
Given the fact that we have the $j_\text{net}$ by introducing Eq. \ref{eq59new} into Eq. \ref{eq58new}, one can simply obtain the ionic current owing to the applied concentration gradient as
\begin{eqnarray}
	I_\text{revED}=e\:\mathcal{A}\:j_\text{net}.
	\label{eq64new}
\end{eqnarray}
Figure \ref{fig7} shows the relation of reverse electro-dialysis current $I_\text{revED}$ to the visco-migration current $I_\text{visco}$ versus the deviation of the right hand-side bulk ionic concentration ($n_\text{b2}$) from the left hand-side bulk concentration ($n_\text{b1}$). To study the impact of the viscosity gradient, bulk ionic concentration, and the solution \textit{p}H on the aforementioned parameters, we consider two viscosity gradients (solid $\eta_1=3.4 \text{(mPas)}, \eta_2=1 \text{(mPas)}$ and dashed lines $\eta_1=3.4 \text{(mPas)}, \eta_2=3 \text{(mPas)}$), two bulk ionic concentration ($n_\text{b}=0.3 \text{mM}$ and $n_\text{b}=100\text{mM}$) and two solution \textit{p}H.
On the one hand, our theoretical model shows that $I_\text{revED}\:I_\text{visco}^\text{-1}$ changes almost linearly with the $n_\text{b2}\:n_\text{b1}^\text{-1}$ for all scenarios (logarithmic $y$-axis). Moreover, it is revealed that by decreasing the viscosity gradient, the impact of diffusion gradient decreases, which results in the domination of the role of the concentration gradient in ionic current. On the other hand, Fig. \ref{fig7} demonstrates that by increasing the bulk ionic concentration or the solution \textit{p}H, the contribution of $I_\text{revED}$ increases in comparison to the $I_\text{visco}$. It is interesting to note that for certain amounts of $n_\text{b2}\:n_\text{b1}^\text{-1}$, $I_\text{revED}\:I_\text{visco}^{-1}$ falls below unity. This implies that for very small applied concentration gradient, the visco-migration ionic current is higher than that from concentration gradient. Generally, when comparing the two contributions, it can be seen that for sufficiently large concentration gradients the reverse ED contribution to the current is orders of magnitude higher than the contribution as a result of the viscosity gradient. 

\section{\label{sec:level5}Conclusions}
In the present contribution, we theoretically investigated the recently discovered ionic transport phenomenon (Wiener and Stein \cite{wiener2018electrokinetic}) which drives ionic species owing to a viscosity gradient along a nanochannel, so-called visco-migration transport phenomenon. Our model has shown that the diffusion gradient can be linearly changed along the nanochannel when we assume the fluids are ideal, while for non-ideal fluids, it showed non-linear behavior as function of the distance from nanochannel entrance. We have shown that the ionic current owing to visco-migration would be a result of nanochannel's surface charge, flux of solvent with higher viscosity, and natural logarithm of lower to higher diffusion coefficient. After benchmarking our ideal and non-ideal binary model with the available experimental data from Wiener and Stein \cite{wiener2018electrokinetic}, we investigated the impact of bulk solution properties (i.e. \textit{p}H and ionic strength) on the visco-migration ionic current by solving the surface complexation models to obtain the relevant surface charge as function of solution bulk properties. Our modeling results suggested that:\par (I) by increasing the solution's bulk ionic strength and \textit{p}H, the absolute value of visco-migration ionic current is increasing, while the increment trend is a function of the applied viscosity gradient. Moreover, we have found that the absolute value of visco-migration ionic current has a maximum point, which can be attributed to the competition of thinning EDL (in contrast to thickening/overlapping EDL) and the increment of surface charge; \par (II) by applying a small ionic concentration gradient at both reservoirs, the impact would be twofold; one can exploit the diffusion gradient and concentration gradient to drive higher amounts of ionic current. For isntance, for $n_{b2}\:n_{b1}^{-1}=0.99$, $n_{b1}=0.3 \: \text{(mM)}$, and $\eta_\text{1}=\text{3.4} \: \text{(mPas)}$ and $\eta_\text{2}=\text{3} \: \text{(mPas)}$ we have greater reverse ED ionic current (order of $4\times10^2$) than the visco-migration ionic current. \par
The theoretical model of this study may pave the way for novel ideas which combine the visco-migration and reverse electro-dialysis for energy conversion applications. Furthermore, this contribution would help researchers to understand how smart selection of solution's properties (i.e. \textit{p}H and bulk ionic strength) can be lead to drive the maximum amount of ionic current from a viscosity gradient.

\section*{Acknowledgment}
AA is very thankful to Prof. Derek Stein and Dr. Benjamin Wiener for fruitful discussions. This work is financially supported by the Eyes High Postdoctoral program at the University of Calgary. AB acknowledges financial support from the Canada First Research Excellence Fund. 
\nocite{*}

\bibliography{apssamp}

\appendix
\section*{Appendix A: Deriving equation for reverse electro-dialysis with viscosity gradient}
Considering the ideal fluid which we already have an explicit solution for the volume fractions of solvents species  (Eq. \ref{eq31new}), one can simply relate the mixed fluid viscosity along the nanochannel to the volume fractions by \cite{RN1854}
\renewcommand{\theequation}{A.\arabic{equation}}
\setcounter{equation}{0}
\begin{equation}
	\eta_\text{12}(x)=\eta_1^{\phi_1(x)}\eta_2^{1-\phi_1(x)}.
	\label{eqA1}
\end{equation}
Having the viscosity along the nanochannel lets us to obtain the diffusivity of the ionic species by employing the Stokes-Einstein equation as
\begin{equation}
	D^\circ(x)=\frac{k_bT}{6\pi\eta_\text{12}(x)r^\circ}
	\label{eqA2}
\end{equation}

Taking $d/dx$ from both sides of Eq. \ref{eqA2} gives
\begin{equation}
	\frac{dD^\circ(x)}{dx}=\frac{-k_bT\mathcal{X}_1\ln\frac{\eta_1}{\eta_2}}{6\pi\eta_\text{12}(x)r^\circ}
	\label{eqA3}
\end{equation} 
where $r^\circ$ denotes the Stokes radius of the ionic species that we assumed approximately identical for both co- and counter-ionic species (Na${^+}$Cl${^-}$) equal with 1.94 \r{{A}}. \par
The flux of ionic species can be described by the Nerns-Planck equation \cite{Daiguji2004} by considering the diffusion and electromigration as
\begin{equation}
	j_i=-\left(D_i\nabla c_i+\frac{z_i}{V_T}D_ic_i\nabla \psi\right).
	\label{eqA4}
\end{equation}
According to the mass conservation law for a steady-state condition without sink or source of ionic species, justifies $\nabla \cdot j_i=0$ which gives rise
\begin{equation}
	0=\nabla D_i \cdot \nabla c_i+D_i\nabla^2 c_i+z_i\nabla  \cdot \left(D_ic_i\nabla\bar{\psi}\right), 
	\label{eqA5}
\end{equation}

where $\bar{\psi}=\psi/V_T$. If we do subtraction over the ionic species for Eq. \ref{eqA5}, we have
\begin{equation}
	0=\nabla D_i \cdot \nabla n_s+D_i\nabla^2 n_s+\nabla  \cdot \left(D_in_d\nabla\bar{\psi}\right) 
	\label{eqA6}
\end{equation}
where $n_d=c_1+c_2$ and $n_s=c_1-c_2$.
Since our nanochannel's wall are impermeable, therefore the flux of ionic species along the $y$-direction must be zero which justifies
\begin{equation}
	j_{i,y}=0 \rightarrow \frac{\partial c_i}{\partial y}+z_ic_i\frac{\partial \bar{\psi}}{\partial y}=0.
	\label{eqA7}
\end{equation}
Now $j_{1,y}-j_{2,y}$ can be written as
\begin{equation}
	0=\frac{\partial n_s}{\partial y}+n_d\frac{\partial \bar{\psi}}{\partial y}.
	\label{eqA8}
\end{equation}
If we take $\partial/\partial y$ from both sides of Eq. \ref{eqA8} and introduce the results into Eq. \ref{eqA6}, we finally have
\begin{equation} \label{eqA9}
	\begin{aligned}
		\Omega \frac{\partial {n_s}}{\partial {x}}+D^{\circ} \frac{{\partial}^2 {n_s}}{\partial {x}^2}+n_d \Omega \frac{\partial {\bar{\psi}}}{\partial {x}} \\
		+D^{\circ} \frac{\partial n_d}{\partial x} \frac{\partial\bar{\psi}}{\partial x}+D^\circ n_d \frac{\partial^2 \bar{\psi}}{\partial x^2}=0.
	\end{aligned}
\end{equation}
where $\Omega=dD^\circ /dx$.
As alluded to what previously we noted, since we are applying very small amounts of concentration gradients and also ignoring the nanochannel entrance effect, therefore it is reasonable to assume that $\partial \bar{\psi}/\partial x =0$, which make Eq. \ref{eqA9} simpler as
\begin{equation}
	\Omega \frac{\partial n_s}{\partial x}+D^\circ \frac{\partial^2 n_s}{\partial x^2}=0.
	\label{eqA10}
\end{equation}
If we take an average in $y-$direction from both sides of Eq. \ref{eqA10} and define $\langle n_s \rangle=\int_{0}^{H/2}n_s/H/2$, then we have
\begin{equation}
	\Omega \frac{d \langle n_s \rangle}{dx}+D^\circ\frac{d^2 \langle n_s \rangle}{dx^2}=0.
	\label{eqA11}
\end{equation}
Eq. \ref{eqA11} cannot be solved analytically since both $\Omega$ (Eq. \ref{eqA3}) and $D^\circ$ (Eq. \ref{eqA2}) are complex functions of $x$. For sake of simplicity, we can utilize Taylor expansion for $D^\circ$ with $\mathcal{O}\left(D^\circ\right)^2$ and maximum error on the order of $10^{-6}$. Consequently, we can re-write Eq. \ref{eqA11} as
\begin{equation}
	B\frac{d \langle n_s \rangle}{dx}+\left(A+Bx\right)\frac{d^2 \langle n_s \rangle}{dx^2}=0.
	\label{eqA12}
\end{equation}
In Eq. \ref{eqA12}, we replace $\Omega$ with the slope of $D^\circ$ which represented by $B$. Eq. \ref{eqA12} can be analytically solved by a proper boundary conditions (see Eqs. \ref{eq59new} to \ref{eq63new}).

\end{document}